\documentstyle[psfig]{elsart}
\begin{document}
\begin{frontmatter}
\title{Specific heat of two--dimensional diluted magnets}
\author{W. Selke  \dag\ , L. N. Shchur \ddag\  and O. A. Vasilyev \ddag }
\address{
\dag Institut f\"ur Theoretische Physik, Technische Hochschule, 
D--52056 Aachen, Germany\\
\ddag Landau Institute for Theoretical Physics, 142432 Chernogolovka, Russia
}
\maketitle
\begin{abstract}
Using Monte Carlo techniques, the two--dimensional site--diluted Ising
model is studied. In particular, properties of the specific
heat, its critical behaviour and the emergence of a non--singular
maximum above the transition temperature at moderate concentration
of defects, are discussed.
\end{abstract}

\begin{keyword}
Two--dimensional Ising model, site--dilution, specific heat,
Monte Carlo simulations\\
PACS: 02.70.Lq, 05.50.+q, 64.60.Ak, 75.40.-s, 75.10.Nr\\
\end{keyword}
\end{frontmatter}

\noindent {\bf 1. Introduction}
 
The effect of randomness on the critical properties of two--dimensional
Ising magnets has attracted much
interest. \cite{dot1,sha,se1} Based on field--theoretical renormalization
group calculations, it has been argued that weak randomness will modify
the critical behaviour of the perfect case marginally. For instance, the
specific heat $C$ is found to diverge on approach to the critical
temperature, $T_c$, in a doubly logarithmic form \cite{dot1,sha}

\begin{equation}
C \propto ln (ln t)  
\end{equation}

\noindent
where $t = |T -T_c|/T_c$ is the reduced temperature. In the pure case, 
$C$ displays the well--known logarithmic singularity.\\

Results of numerical studies, using Monte Carlo methods and 
finite--size transfer matrix techniques, on the two--dimensional
{\it bond}--diluted Ising model, even at moderate or rather strong
dilution, are in accordance with the field--theoretical
predictions \cite{wang,que,reis,tal,ts2,stauff1,igloi}, albeit some of the
data leave room for alternate interpretations.-- Attention may be also
drawn to related recent simulations on two--dimensional Potts and
Ashkin--Teller models. \cite{wise,berche}\\

On the other hand, simulations on the two--dimensional {\it site}--diluted 
Ising model seemed to indicate that the critical peak in $C$, present
at weak dilution, disappears at moderate dilution,
well below the percolation threshold \cite{stauff}, with
a broad maximum above the critical
temperature. \cite{stoll,huber,vel,heuer,kim} Certainly, a finite
value of the specific heat at $T_c$ would invalidate (1).
However, recent Monte Carlo work \cite{ball} suggests
that the field--theoretical results may be correct also in the case of moderate
site--dilution. Indeed, $C(T_c)$ is found to increase with the
linear size, $L$, of the system like $ln (ln L)$. Accordingly, quite
large system sizes might be required to
monitor a peak in $C$ near $T_c$.\\

In any event, a careful and systematic investigation on the emergence of the
non--critical broad maximum in the specific heat and on the fate of
the critical peak in $C$ with increasing dilution
is much needed. In this paper, we shall present
results of such a study.\\

\noindent {\bf 2. Model, method and results}

We consider a square lattice with sites $i$. A randomly
chosen fraction $1-p$ of the
sites are assumed to be occupied by Ising spins, $S_i = \pm 1$, coupled
by ferromagnetic nearest--neighbour interactions, $J$. The remaining
fraction, $p$, of sites is 'empty' (or occupied by non--magnetic ions).
Accordingly, $p =0$ refers to the perfect case. The
transition temperature $T_c(p)$ is lowered by increasing the concentration
of defects, $p$. It vanishes at and above the percolation
threshold, $p_c \approx 0.40725$ \cite{stauff,stinch}, see Fig. 1.\\

\begin{figure}
\centerline{\psfig{figure=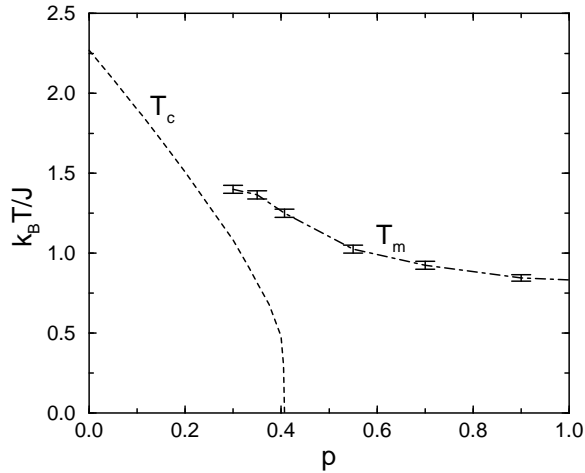,width=8.5cm,angle=270}}
\caption{Phase diagram of the two--dimensional site diluted Ising model
with the transition line $T_c$ and the location
of the non--critical maximum $T_m$ of the specific heat,
as obtained in previous and the present simulations.}
\label{fig1}
\end{figure}

The simulations were done using the Wolff cluster--flip algorithm
and, for small lattices or at temperatures well
above $T_c$, the Metropolis single--spin--flip
method. We considered quadratic lattices of $L \times L$ sites with full
periodic boundary conditions; $L$ ranging from 8 to 256. The
dilution $p$ varied in between 0 and 0.9. Apart from the
energy, $E$, and the specific heat per spin (which we computed both from
the energy fluctuations and the temperature derivative of the
energy), $C$, we recorded
the absolute value of the magnetization $|m|$ 
and the susceptibility, $\chi$ (as well
as other quantities related to properties of the Wolff
clusters \cite{vasil}). We monitored the behaviour of single
realizations of the site--randomness and of ensembles of $N$ 
realizations (with $N$ going up to 1000). For each realization, we
averaged over up to $10^7$ clusters
(or up to several $10^6$ Monte Carlo steps per site), after
equilibration.\\

To set the frame for the following discussion, the
phase diagram is depicted in Fig. 1, based on previous
large--scale simulations \cite{heuer,kim,ball}, which we checked
and augmented. The phase transition line, $T_c(p)$, is observed
to decrease almost
linearly with dilution up to about $p \approx 0.3$, and then
it bends over to vanish at the percolation threshold, i.e. $T_c(p_c) =0$. In 
adddition, the location of the non--critical
maximum in the specific heat, $T_m(p)$, is shown in Fig. 1 (its location
does not depend strongly on system size, see below).\\
 
For illustration, examples of the temperature dependence of the specific
heat, $C(T)$, at various concentrations of defects $p$, are
shown in Fig. 2, with linear dimension $L = 64$,
for single realizations of the site--dilution. At weak
randomness, one observes a pronounced peak around the critical
temperature, with the height decreasing rapidly with increasing
dilution. At moderate
dilution ($p = 0.26$, in the figure), the specific heat displays at 
temperatures above $T_c$ a shoulder, eventually 
turning over into the non-critical
maximum, at $T_m(p)$, which persists at defect concentrations above the
percolation threshold up to arbitrarily large dilution.
The height of the non--critical maximum decreases
only mildly for stronger dilution.\\

\begin{figure}
\centerline{\psfig{figure=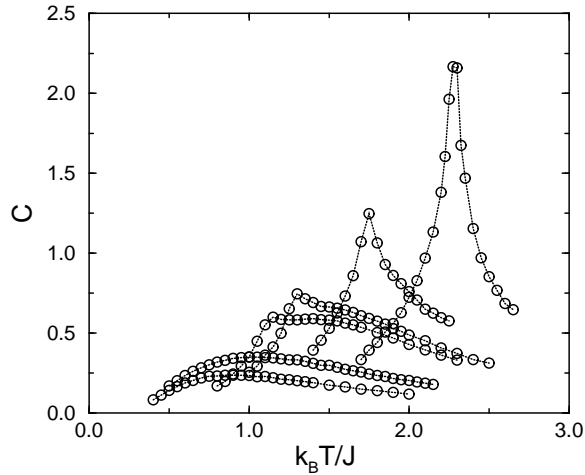,width=8.5cm,angle=270}}
\caption{Specific heat, $C$, versus temperature, $k_b T/J$, for
single realizations with $L$= 64 at various concentrations of defects,
$p$= 0.15, 0.26, 0.3, 0.55, and 0.7; see also Fig. 1.} 
\label{fig2}
\end{figure}

The region of moderate dilution, say, $0.25 < p < 0.35$, deserves
special attention, in order to, possibly, disentangle the critical
peak in $C$ from the non--critical maximum above $T_c$.
In particular, the kind of the realization, i.e. of  
the distribution of the spins on the lattice, and the
system size are important. We did an extensive
study of this region by considering mostly $p$= 0.26, 0.30, and 0.35,
monitoring a wide range of realizations and lattice sizes.\\  

Note that very good statistics is needed to establish unambiguously
especially the subtle features in the shape of $C$ in that 
region. The accuracy of the data has been checked, for instance, by 
looking for consistency in the averages for $C$ as
obtained from the temperature derivative and from the fluctuations
of the energy. Eventually, we took into account up to $10^7$
clusters, close to $T_c$, or several $10^6$ Monte Carlo steps
per site, using the Wolff or Metropolis algorithm.
We checked and confirmed, that our results are independent
of the type of the
simulational algorithm (the
two methods are complementary, with the Wolff algorithm
being more powerful at low temperatures and close to
criticality for large systems).
We also used different random number
generators to avoid possible dangerous correlations arising
from an unfortunate choice of that generator \cite{seshta} (indeed,
for the dilute model, both linear congruential and shift register
random number generators are suitable). Note that previous
simulational results, when based on significantly shorter
Monte Carlo runs, should be viewed with much care.\\

Fig. 3 summarizes our findings on the specific heat $C$ at $p= 0.3$, with
$L$ ranging from 8 to 256, where the number of maximal realizations $N$
for the various lattices is decreasing from 1000 to 4. The
statistical errors for each
realization are very small. However, deviations between different
samples, fixing the number of sites and spins, may be large, especially
for small lattices (say, up to $L$= 64;
at $L= 64$, some of the single realizations still show
only the broad maximum above $T_c$, while others
exhibit an additional, albeit weak, peak at $T_c$). Therefore, the
total bars stem mostly from the ensemble
averaging. By adjusting the number of samples, $N$, to the size
of the lattice, $L$, the resulting errors were always smaller
than the size of the symbols in Fig. 3. As seen from that figure, the
non--critical maximum in $C$ and the peak close to $T_c$ can
be easily discriminated by simulating sufficiently large systems (which
had not been noted before). The shape of the non-critical
maximum becomes independent of the lattice size for 
sufficiently large systems. It
may be interesting to note that the appearance of a rather narrow
peak at $T_c$ had been suggested before by one of us \cite{selkex}, discussing
possible similarities to analytic findings on other random
models \cite{zitt1}. Doing standard finite--size analyses \cite{fisher},
 we estimated the critical temperature $T_c$ from the location of
the turning point in $|m|$ and from the critical peak in $C$ to be
$k_B T/J= 1.084 \pm 0.001$ (being slighly higher than a previous
estimate \cite{heuer}). We found the shift of
the deviation of the location of the anomalies from $T_c$ to be nearly
proportional to $1/L$ for sufficiently large $L$ ($\ge 64$).  
The specific heat at $T_c$ for finite lattices, $C(T_c,L)$, was 
then obtained by interpolation between
close--by data of $C(T)$, leading to increased error bars.\\

\begin{figure}
\centerline{\psfig{figure=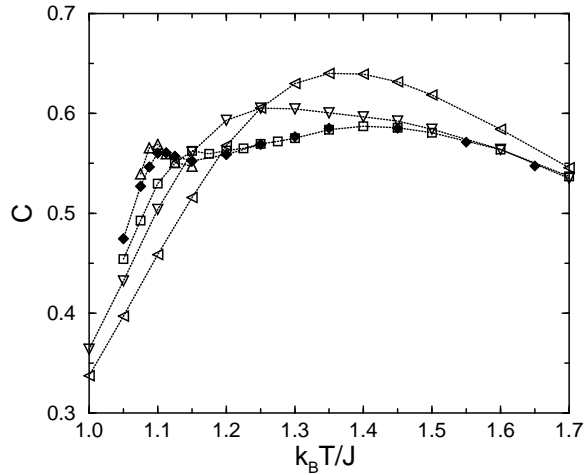,width=8.5cm,angle=270}}
\caption{Specific heat vs. temperature at fixed dilution, $p =0.30$, for
different system sizes, $L$= 16 (triangle left), 32 (triangle down),
 64 (square), 128 (diamond) and 256 (triangle up), averaging over up to
200, 48, 15, 8, and 4 realizations, respectively. Error bars
are smaller than the sizes of the symbols.}
\label{fig3}
\end{figure}

As depicted in Fig. 4a, C($T_c, L)$ seems to approach, for $L \ge 32$,
a doubly logarithmic form, $C = C_0 ln (ln L) + C_1$, in accordance with
the field--theoretical prediction for weak randomness. The prefactor
$C_0$ is quite small, $C_0 \approx 0.17$ (with $C_1 \approx 0.27$), similarly
to findings in the bond--diluted case
for moderate dilution. \cite{wang} Consistently, $C(T_c, L)$ seems to 
increase more slowly than logarithmically for sufficiently
large lattices, while a logarithmic increase is
conceivable for smaller lattices, see Figs. 4a and 4b. The
crossover to the dilution dominated critical regime 
may be described by casting $C(T_c,L)$ in the
form $C = C_0 ln (b + ln L) + C_1$, as had
been obtained before for the
bond--diluted Ising model. \cite{wang} We confirmed that the
plots are insensitive towards the exact determination
of $T_c$, accepting the accuracy of our
estimate ($k_B T_c/J= 1.084 \pm 0.001$). Of course, the
situation in the bond--diluted Ising model, where
$T_c$ is known exactly, is more convenient. \cite{wang}\\

\begin{figure}
\centerline{\psfig{figure=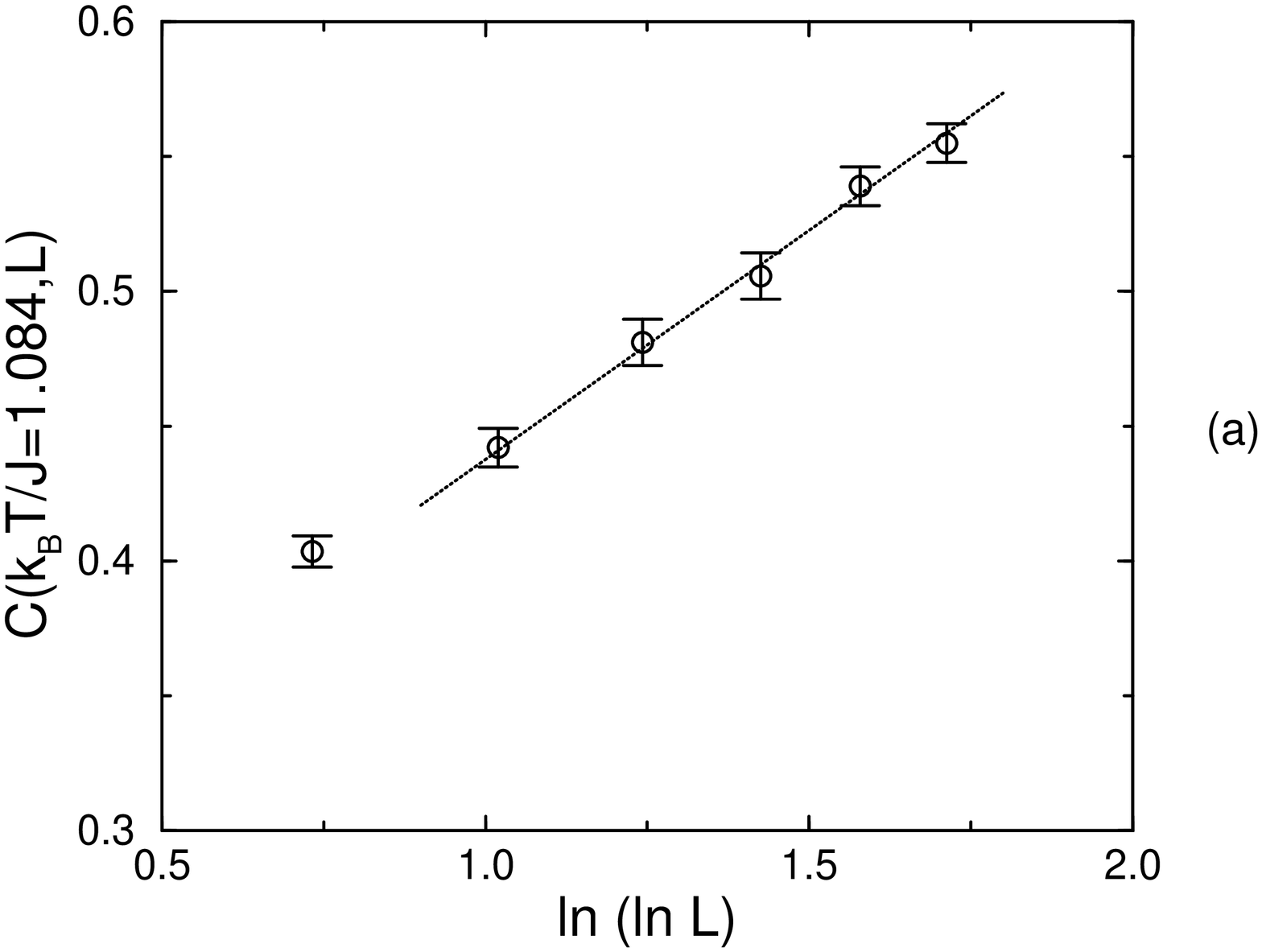,width=8.5cm,angle=270}}
\centerline{\psfig{figure=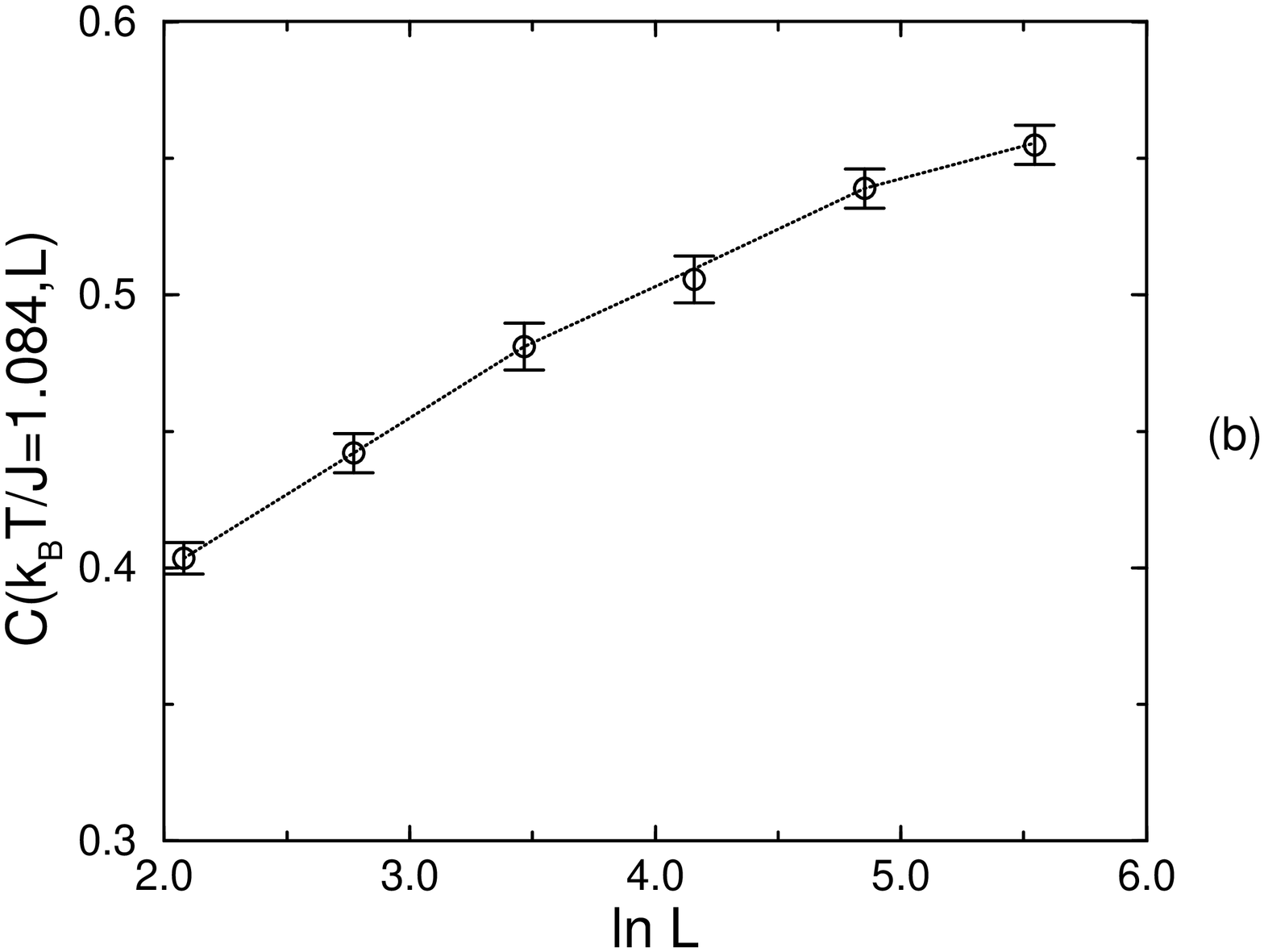,width=8.5cm,angle=270}}
\caption{Specific heat $C$ vs. lattice size $L$ ,
using a doubly logarithmic (a) and a logarithmic (b) scale
for $L$, at the estimated critical temperature $k_B T_c = 1.084 J$ for
$p$= 0.30. Lines are guides to
the eye.}
\label{fig4}
\end{figure}

At $p$= 0.26, one observes a broad shoulder in $C$ above $T_c$, which
one may associate with the non--critical maximum, superimposed and
masked by critical fluctuations. At $p$= 0.35, the
specific heat displays a broad maximum well above $T_c$, compare to
Figs. 1 and 2. Around $T_c$ (estimated from $\chi$ and $d|m|/dt$), $C$
increases with the lattice size. However, even
for $L= 256$, there is only a pronounced shoulder in $C$ close to
$T_c$, but no peak. Presumably, much larger systems are required
to detect that supposedly narrow peak
at $T_c$. We shall discuss this issue
from a microscopic point of view in the following.\\

In a microscopic description, the neighbouring lattice sites occupied
by spins form distinct clusters. Certainly, spins in different
clusters do not interact with each other, and thermodynamic quantities
are obtained by summing over the contributions of separate clusters. For 
instance, the specific heat may be written as
$C= \sum  C_k$, summing  
over all clusters, $k$.\\

At $p < p_c$, in the thermodynamic limit $L \longrightarrow \infty$, there
exists one cluster with infinitely many spins covering a non--zero
fraction of the lattice sites. That cluster, corresponding to the largest
cluster in finite systems, is expected to carry the critical
properties, and the other remaining clusters do not lead to non--analytic
thermal averages. Of course, the
weight of the contributions of the remaining clusters to the non--singular
features of many thermal quantities is expected to
grow as the dilution increases. It may be therefore tempting to attribute
the non--critical maximum in $C$ to those clusters. However, this 
argument is not correct.\\

In particular, for defect concentrations $p$ of about 0.3, we determined the
contributions to $C$ due to the largest and the other
clusters separately. We found that the
energy fluctuations of the other clusters are far too small to account
for the shoulder and the non--critical maximum
in $C$ above $T_c$. Actually, most of the
finite clusters consist of single spins, with vanishing specific heat. On 
the other hand, the largest cluster has a rather ramified structure. There
are many weakly connected subclusters of various sizes, which may
act essentially like separate clusters. \cite{stauff} Those
subclusters may flip
completely near $T_c$, with a small
change of energy resulting from the spins at the links between
them. Accordingly, the contributions of such 
excitations to the specific heat (i.e. the temperature derivative
of the energy) may be still small. These excitations, however, affect
strongly the magnetization and its fluctuations, i. e. the
susceptibility, leading to pronounced peaks in $d|m|/dt$ and  
in $\chi$ near $T_c$. Presumably most importantly, the number of
perimeter spins, i.e. spins with neighbouring empty sites, is quite
large \cite{stauff}, reflecting the loose
structure of the largest cluster. The perimeter spins, which
have a reduced coordination number, are thermally excitable at
a characteristic temperature, leading to near--by local disordering
and a pronounced change in energy in that temperature region. Therefore, they 
may contribute significantly to the shoulder and, upon further
dilution, to the non--critical maximum (note that this feature
is valid and specific for site--dilution. Indeed, for bond--diluted
Ising models, $C$ shows no anomalies above the critical point \cite{in}). 
The critical behaviour of $C$, in turn, is expected to be due
to the rather compact
backbone of the largest cluster \cite{stauff}. The size of the backbone scales
with the lattice size, $L^2$, at $p < p_c$, and it may eventually
give rise to a divergent peak in $C$.\\

Increasing the dilution, with $0.3 < p < p_c$, a growing number of spins will 
no longer belong to the largest cluster, which, itself, will
become even more ramified. As a result the non--critical
energy fluctuations will be enhanced, and the, supposedly, critical
contributions of the compact backbone of the largest cluster
become more and more suppressed, requiring larger and larger
system sizes to detect the possibly singular critical behaviour of $C$.\\

At $p > p_c$, in the thermodynamic limit, there is no cluster of spins
covering a non--zero fraction of the lattice (the largest cluster
is believed to grow with system size like $ln L$ \cite{stauff}, while
the number of lattice sites increases with $L^2$). 
As a consequence, e.g., the infinite cluster cannot stabilize the
ordered phase at non-zero temperatures, and
$T_c(p) = 0$. Obviously, the remaining finite clusters contribute more
significantly to, e.g., the specific heat at stronger dilution. Their
average size shrinks with rising dilution, leading to the shift of
the broad maximum in $C$ towards lower temperatures, $T_m$. Actually, the
smallest temperature would be reached if there would be predominantly
clusters consisting of one or two spins. The maximum of $C$ for clusters
of two spins is readily calculated to be $T_m= 0.8335... J/k_B$. Indeed, this
temperature seems to be approached when $p \longrightarrow 1$, see
Fig.1.\\

\noindent {\bf 3. Summary}

Using Monte Carlo techniques, the specific heat $C$ of the 
site--diluted two--dimensional Ising model has been studied
as a function of temperature, lattice size, and degree of
dilution, considering single realizations of the dilution as
well as ensembles of samples.\\
 
Upon increasing the dilution, we observe the systematic
evolution of a non--critical maximum above the critical point. In
turn, eventually, rather large system sizes are then
needed to detect the narrow critical peak at $T_c$. Indeed, we showed
unambiguously the existence of the corresponding two--maxima structure
in $C$ at moderate dilution, $p=0.3$, for lattices with linear
dimension $L \ge 128$. That aspect had not been noted
before. The critical behaviour of the peak
at $T_c$ seems to be, at $p=0.3$, in
accordance with the field--theoretical predictions. In particular, 
the Monte Carlo data are compatible with a crossover, for
sufficiently large lattices, to the 
dilution dominated regime where $C(T_c,L)$ increases
in a doubly logarithmic form with $L$.\\

A microscopic picture, based on the geometric aspects of
the clusters of spins (perimeter and backbone
spins of the largest cluster, weakly
coupled subclusters and separate small clusters), is
proposed allowing to describe qualitatively, and partly
even quantitatively, the properties
of the specific heat observed in the simulations.\\ 
 
\noindent {\bf Acknowledgements}
  
We should like to thank D. Stauffer for a very useful discussion, and
INTAS (through grant 93-211) as well as RFBR for financial support.\\

\end{document}